\documentclass[aps,pre,twocolumn,superscriptaddress,floatfix,times]{revtex4}
\usepackage{graphicx}
\usepackage[dvips,unicode,colorlinks,linkcolor=blue,citecolor=blue,urlcolor=blue]{hyperref}
\begin{document}

\title
{Suppression of thermal conductivity in graphene nanoribbons
with rough edges}

\author{Alexander V. Savin}

\affiliation{Semenov Institute of Chemical Physics, Russian Academy
of Sciences, Moscow 119991, Russia}
\affiliation{Nonlinear Physics Center, Research School of Physics and Engineering,
Australian National University, Canberra, ACT 0200, Australia}

\author{Yuri S. Kivshar}
\affiliation{Nonlinear Physics Center, Research School of Physics and Engineering,
Australian National University, Canberra, ACT 0200, Australia}

\author{Bambi Hu}
\affiliation{
University of Houston,
Department of Physics, University of Houston, Houston,
TX 77204-5005, USA}
\affiliation{
Centre for Nonlinear Studies, and The Beijing-Hong Kong-Singapore Joint Centre
for Nonlinear and Complex Systems (Hong Kong), Hong Kong Baptist University,
Kowloon Tong, Hong Kong, China}

\begin{abstract}
We analyze numerically the thermal conductivity of carbon nanoribbons with
ideal and rough edges. We demonstrate that edge disorder can lead to a
suppression of thermal conductivity by several orders of magnitude. This effect is
associated with the edge-induced Anderson localization and suppression of the phonon
transport, and it becomes more pronounced for longer nanoribbons and low temperatures.
\end{abstract}


\maketitle

\section{Introduction}
The study of remarkable properties of graphite structures is one of the hot topics
of nanoscience~\cite{Kats2007}. Graphene nanoribbons (GNRs) are effectively
low-dimensional structures
similar to carbon nanotubes, but their main feature is the presence of edges. Due to the edges,
graphene nanoribbons can demonstrate many novel properties driven by their geometry,
depending on
their width and helicity. A majority of the current studies of graphene nanoribbons
are devoted to the analysis of  their electronic and magnetic properties modified by
the presence of edges, including the existence of the localized edge modes~\cite{Lee2009,Engelund},
which are an analog of surface states in the two-dimensional geometry. The edge can support
localized vibrational states in both linear and nonlinear regimes~\cite{sk10prb,sk10epl}.

The effect of the edge disorder on the electronic transport of graphene nanoribbons
has been discussed in several papers (see, e.g., Refs. ~\cite{ref1,ref2,ref3}).  It was found
that already very modest edge disorder is sufficient to induce the conduction
energy gap in the
otherwise metallic nanoribbons and to lift any difference in the conductance between nanoribbons
of different edge geometry, suggesting that this type of disorder can be very important for
altering other fundamental characteristics of GHRs.

In addition to electronic properties, the thermal properties of graphene are also
of both fundamental and practical importance. Several experiments~\cite{Ghosh,Balandin}
have demonstrated that graphene has a superior thermal conductivity, likely underlying
the high thermal conductivity known in carbon nanotubes~\cite{Pop}.
This opens numerous possibilities for using graphene nanostructures in nanoscale
thermal circuit management.

Recent experiments demonstrated that thermal conductivity of
silicon nanowires can be dramatically reduced
by surface roughness~\cite{Hochbaum,Boukai}. This results has been
confirmed theoretically in the framework
of a simplest phenomenological model of quasi-one-dimensional crystal
that demonstrates the reduction of thermal conductivity
due to roughness-induced disorder~\cite{ks09epl}.
Molecular dynamics simulations~\cite{Hu} demonstrated that
thermal conductivity of GRNs depends on the edge chirality and can be affected by defects.
Therefore, we wonder if the edge disorder of GNRs can modify substantially their
thermal conductivity, similar to the case of
silicon nanowires.

In this Article, we study the thermal conductivity of isolated graphene nanoribbons
with ideal and rough edges.
By employing a direct modeling of heat transfer by means of the
molecular-dynamics simulations, we demonstrate that
the thermal conductivity grows monotonically with the GNR
length as a power-law function. In contrast,
rough edges of the nanoribbon can reduce the thermal conductivity by several
orders of magnitude.  This effect is enhanced
for longer GNRs and for lower temperatures, and it corresponds
to dramatic suppression of phonon transport
solely by the edge disorder.  It means that nanoribbons  with ideal edges can play a role of
highly efficient conductors in nanocircuits, whereas the rough edges will transform
them into efficient thermal resistors.

\section{Model}
\begin{figure}[tb]
\includegraphics[angle=0, width=1.\linewidth]{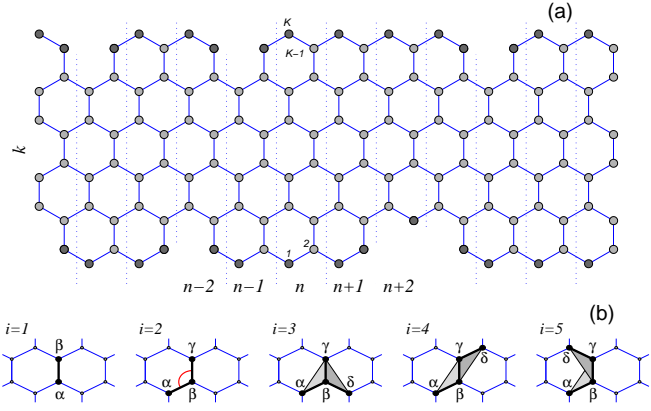}
\caption{
(a) Schematic view of a zigzag nanoribbon with rough edges
and atom numbering. The edge atoms are shown as filled circles.
Dotted lines separate the elementary cells of the nanoribbon.
$K$ is the number of atoms in the elementary cell.
(b) Configurations of an ideal structure containing up to $i$-th nearest-neighbor
interactions for $i=1,...,5$.
}
\label{fig1}
\end{figure}

We model a graphene nanoribbon as a planar strip of graphite,
with the properties depending on
the stripe width and chirality. The structure of  the zigzag nanoribbon can be presented
as a longitudinal repetition of the
elementary cell composed $K$ atoms (the even number $K\ge 4$).
We use atom numbering shown in Fig. \ref{fig1}(a). In this case, each carbon
atoms has a two-component index $\alpha=(n,k)$, where $n=0,\pm1,\pm2,...$ stands for the
number of the elementary cells, and $k=1,2,...,K$ stands for the number atoms in the cell.

Each elementary cell of the zigzag nanoribbon has two edge atoms. In Fig. \ref{fig1}(a),
we show these edge atoms as filled circles. We consider a hydrogen-terminated
nanoribbon, where edge atoms correspond to the molecular group CH. We consider such a group
as a single effective particle at the location of the carbon atom. Therefore, in our model
of graphene nanoribbon we take the mass of atoms inside the strip as $M_0=12m_p$, and for the
edge atoms we consider a large mass $M_1=13m_p$ (where $m_p=1.6603\cdot 10^{-27}$kg is the
proton mass).
%---------------------------- Fig. 1 ------------------------------------
%---------------------------- Fig. 1 ------------------------------------

\begin{figure}[tbhp]
\includegraphics[angle=0, width=1.\linewidth]{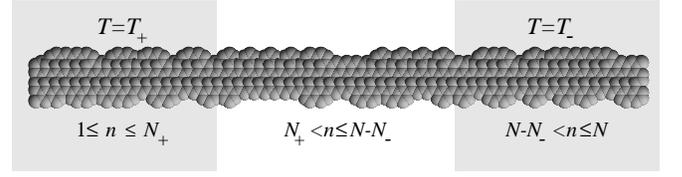}
\caption{
Example of a zigzag nanoribbon with rough edges
with $N$ longitudinal segments. First left $N_+$ segments are attached
to the $T=T_+$ thermostat and the last right $N_-$ segments are
attached to the $T=T_-$ thermostat. Number of atoms in the elementary
cell $K=12$, density
of the nanoribbon with rough edges $d=0.87$ (probability $p=0.5$).
\label{fig2}
}
\end{figure}

To model two rough edges we randomly delete some atoms with second index $k=1$ and $k=K$.
Let $0\le p\le 1$ be the probability of atom removal. As a  result of the random atom removal
from the edge layers, some atoms at the edges will have only one covalent bond C--C and
should be deleted as well. After this operation,
the edge become rough, as shown in Fig. \ref{fig2}.
Here all edge atoms participate only in two valent bonds C--C.
We characterize the degree of roughness by the parameter $d=N_a/N_b$,
where $N_a$ is the number of atoms in an ideal nanoribbon, and $N_b$
is the number of atoms remaining in the edge-disordered nanoribbon after
removing some of the edge atoms.
Parameter $d$ characterizes the density of the edge-disordered nanoribbon
in comparison with the ideal case.
When the probability $p$ for removal of an edge atom is $p=0$,
we have $d=1$, and $d$ decays for larger values of
the density $p$, so that for $p=1$ (when all atoms with the second
index $k=1,2$ and $k=K-1,K$ are removed),
it takes the minimum value $d=(K-4)/K$ (for $p=1$ we have again an
ideal ribbon but for a smaller width, with
$K-4$ atoms in an elementary cell).  For $K=12$ and probability $p=0.5$
the density is $d=0.87$: this
nanoribbon is shown in Fig. \ref{fig2}.
%---------------------------- Fig. 2 ------------------------------------
%---------------------------- Fig. 2 ------------------------------------

To describe the dynamics of both ideal and disordered nanoribbons, we present the
system Hamiltonian in the form,
\begin{equation}
H = \sum_{n=-\infty}^{+\infty}\sum_{k=1}^{K_n} \left[\frac12 M_{(n,k)}
(\dot{\bf u}_{(n,k)},\dot{\bf u}_{(n,k)})+P_{(n,k)}\right],
\label{f1}
\end{equation}
where $K-4\le K_n\le K$ is the number of atoms in the $n$-th elementary cell,
$M_\alpha$ is the mass of the hydrogen atom with the index $\alpha=(n,k)$
(for internal atoms we take $M_\alpha=M_0$, whereas for the edge atoms we take a larger mass,
$M_\alpha=M_1>M_0$), ${\bf u}_\alpha=(x_\alpha(t),y_\alpha(t),z_\alpha(t))$
is the radius-vector of the carbon atom with the index $\alpha$ at the moment $t$.
The term $P_\alpha$ describes the interaction of the atom with the index $\alpha=(n,k)$
with its neighboring atoms.
The potential depends on variations of bond length, bond angles, and dihedral angles
between the planes formed by three neighboring carbon atoms, and it can be written in the form
\begin{equation}
P=\sum_{\Omega_1} U_1+\sum_{\Omega_2} U_2+\sum_{\Omega_3} U_3+\sum_{\Omega_4} U_4
+\sum_{\Omega_5} U_5,
\label{f2}
\end{equation}
where $\Omega_i$, with $i=1,2,3,4,5$ stand for the sets of configurations including up to
nearest-neighbor interactions. Owing to a large redundancy, the sets only
need to contain configurations of the atoms shown in Fig. \ref{fig1}(b), including their
rotated and mirrored versions.

The potential $U_1({\bf u}_\alpha,{\bf u}_\beta)$ describes the deformation energy
due to a direct interaction between pairs of atoms with the indices
$\alpha$ and $\beta$, as shown in Fig. \ref{fig1}(b).
The potential $U_2({\bf u}_\alpha,{\bf u}_\beta,{\bf u}_\gamma)$
describes the deformation energy of the angle between the valent bonds
${\bf u}_\alpha{\bf u}_\beta$ and  ${\bf u}_\beta{\bf u}_\gamma$.
Potentials $U_{i}({\bf u}_\alpha,{\bf u}_\beta,{\bf u}_\gamma,{\bf u}_\delta)$, $i=3$, 4, 5,
describes the deformation energy associated with a change of the effective angle between
the planes ${\bf u}_\alpha,{\bf u}_\beta,{\bf u}_\gamma$ and
${\bf u}_\beta,{\bf u}_\gamma,{\bf u}_\delta$.

We use the potentials employed in the modeling of the dynamics of large polymer
macromolecules~\cite{Noid,Savin03} for the valent bond coupling,
\begin{equation}
U_1({\bf u}_1,{\bf u}_2)=\epsilon_1
\{\exp[-\alpha_0(\rho-\rho_0)]-1\}^2,~~~\rho=|{\bf u}_2-{\bf u}_1|,
\label{f3}
\end{equation}
where $\epsilon_1=4.9632$~eV is the energy of the valent bond and $\rho_0=1.418$~\AA~
is the equilibrium length of the bond;
the potential of the valent angle
\begin{eqnarray}
U_2({\bf u}_1,{\bf u}_2,{\bf u}_3)=\epsilon_2(\cos\varphi-\cos\varphi_0)^2,
\label{f4}\\
\cos\varphi=({\bf u}_3-{\bf u}_2,{\bf u}_1-{\bf u}_2)/
(|{\bf u}_3-{\bf u}_2|\cdot |{\bf u}_2-{\bf u}_1|),
\nonumber
\end{eqnarray}
so that the equilibrium value of the angle is defined as $\cos\varphi_0=\cos(2\pi/3)=-1/2$;
the potential of the torsion angle
\begin{eqnarray}
\label{f5}
U_i({\bf u}_1,{\bf u}_2,{\bf u}_3,{\bf u}_4)=\epsilon_i(1-z_i\cos\phi),\\
\cos\phi=({\bf v}_1,{\bf v}_2)/(|{\bf v}_1|\cdot |{\bf v}_2|),\nonumber \\
{\bf v}_1=({\bf u}_2-{\bf u}_1)\times ({\bf u}_3-{\bf u}_2), \nonumber \\
{\bf v}_2=({\bf u}_3-{\bf u}_2)\times ({\bf u}_3-{\bf u}_4), \nonumber
\end{eqnarray}
where the sign $z_i=1$ for the indices $i=3,4$ (equilibrium value of the torsional angle $\phi_0=0$)
and $z_i=-1$ for the index $i=5$ ($\phi_0=\pi$).

The specific values of the parameters are $\alpha_0=1.7889$~\AA$^{-1}$,
$\epsilon_2=1.3143$ eV, and $\epsilon_3=0.499$ eV, and they are found from the frequency
spectrum of small-amplitude oscillations of a sheet of graphite~\cite{Savin08}.
According to the results of Ref.~\cite{Gunlycke} the energy $\epsilon_4$ is close to
the energy $\epsilon_3$, whereas  $\epsilon_5\ll \epsilon_4$
($|\epsilon_5/\epsilon_4|<1/20$). Therefore, in what follows we use the values
$\epsilon_4=\epsilon_3=0.499$ eV and assume $\epsilon_5=0$, the latter means that we
omit the last term in the sum (\ref{f2}).

\section{Methods}
In order to model the heat transport, we consider the nanoribbon of a finite length
with two ends places in thermostats kept at different temperatures, as shown
schematically in Fig. \ref{fig2}. In order to calculate numerically the coefficient of
thermal conductivity, we should calculate the heat flux at any
cross-section of the nanoribbon. Therefore, first we obtain the formula
for calculating the longitudinal local heat flux.

We define the $3K_n$-dimensional coordinate vector
${\bf u}_{n}=\{x_{n,k},y_{n,k},z_{n,k}\}_{k=1}^{K_n}$
which determines the atom coordinates of an elementary cell $n$, and then write
the Hamiltonian (\ref{f1}) in the form,
\begin{equation}
H=\sum_nh_n=\sum_{n}[\frac12({\bf M}_n\dot{\bf u}_{n},\dot{\bf u}_{n})
+P_n({\bf u}_{n-1},{\bf u}_{n},{\bf u}_{n+1})],
\label{f6}
\end{equation}
where the first term describes the kinetic energy of the atoms
(${\bf M}_n$ is diagonal mass matrix of the $n$-th elementary cell),
and the second term describes
the interaction between the atoms in the cell and with the atoms of neighboring cells.

Hamiltonian (\ref{f6}) generates the system of equations of motion,
\begin{equation}
-{\bf M}_n\ddot{\bf u}_{n}={\bf F}_n={\bf P}_{1,n+1}+{\bf P}_{2,n}+{\bf P}_{3,n-1},
\label{f7}
\end{equation}
where the function
${\bf P}_{i,n}={\bf P}_i({\bf u}_{n-1},{\bf u}_{n},{\bf u}_{n+1})$,
${\bf P}_i=\partial P({\bf u}_1,{\bf u}_2,{\bf u}_3)/\partial{\bf u}_i$,
$i=1,2,3$.

Local heat flux through the $n$-th cross-section, $j_n$, determines a local longitudinal
energy density $h_n$ by means of a discrete continuity equation,
$\dot{h}_n=j_n-j_{n+1}$.
Using the energy density from Eq.~(\ref{f6}) and the motion equations (\ref{f7}),
we obtain the general expression for the energy flux through the $n$-th cross-section
of the nanotube,
$
j_n=({\bf P}_{1,n},\dot{\bf u}_{n-1})-({\bf P}_{3,n-1},\dot{\bf u}_n).
$

For a direct modeling of the heat transfer along the nanoribbon, we consider a nanoribbon of
a fixed length $(N-1)h$ with fixed ends. We place the first $N_+=40$ segments into the
Langevin thermostat at $T_+=310$K, and the last $N_-=40$ segments, into the thermostat at
$T_-$=290K -- see Fig. \ref{fig2}. As a result, for modeling of the thermal conductivity we
need integrating numerically the following system of equations,
\begin{eqnarray}
{\bf M}_n\ddot{\bf u}_n &=&-{\bf F}_n-\Gamma{\bf M}_n\dot{\bf u}_n+\Xi_n^+,~~\mbox{for}~~n=2,...,N_+,
\nonumber\\
{\bf M}_n\ddot{\bf u}_n &=&-{\bf F}_n,~~\mbox{for}~~n=N_++1,...,N-N_-,\label{f8}\\
{\bf M}_n\ddot{\bf u}_n &=&-{\bf F}_n-\Gamma{\bf M}_n\dot{\bf u}_n+\Xi_n^-,\nonumber\\
&~&\mbox{for}~~n=N-N_-+1,...,N,
\nonumber
\end{eqnarray}
where $\Gamma=1/t_r$ is the damping coefficient (relaxation time $t_r=0.1$ ps), and
$$
\Xi_n^\pm=(\xi_{1,1},\xi_{1,2},\xi_{1,3},...,\xi_{K_n,1},\xi_{K_n,2},\xi_{K_n,3})
$$
is $12K_n$-dimensional
vector of normally distributed random forces normalized by conditions
$$
\langle\xi_{n,i}^\pm(t_1)\xi_{l,j}^\pm(t_2)\rangle=2M_{n,i}k_BT_\pm\delta_{nl}\delta_{ij}\delta(t_1-t_2).
$$
Details of the numerical procedure for modeling of thermal systems can be found elsewhere~\cite{Savin09}).

We select the initial conditions for system (\ref{f8})
corresponding to the ground state of the
nanoribbon, and solve the equations of motion numerically tracing the transition
to the regime with a stationary heat flux. At the inner part of the nanotube
($N_+<n\le N-N_-$), we observe
the formation of a temperature gradient corresponding to a constant flux.
Distribution of the average values of temperature and heat flux
along the nanotube can be found in the form,
\begin{eqnarray*}
T_n=\lim_{t\rightarrow\infty} \frac{1}{3K_nk_Bt}\int_0^t
({\bf M}_n\dot{\bf u}_n(\tau),\dot{\bf u}_n(\tau))d\tau,\\
J_n=\lim_{t\rightarrow\infty}\frac{h}{t}\int_0^t j_n(\tau)d\tau,
\end{eqnarray*}
where $k_B$ is the Boltzmann constant. For nanoribbons with rough edges
we make the averaging
not only in  time but also on 240 independent realizations of the roughness.

\begin{figure}[tbh]
\includegraphics[angle=0, width=1.\linewidth]{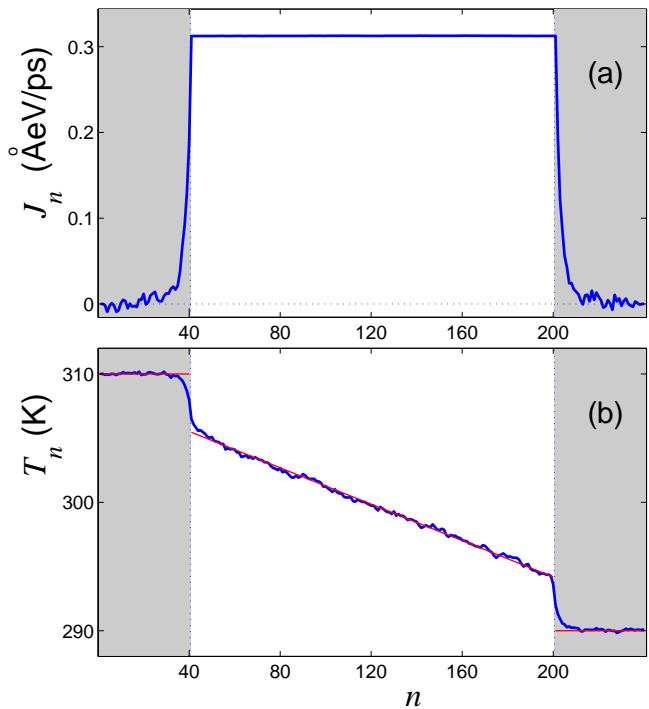}
\caption{
(Color online) Distribution of (a) local heat flux $J_n$  and (b) local average temperature
$T_n$ along ideal zigzag nanoribbon with $K=8$.
Length of the nanoribbon is
$L=(N-1)h= 58.8$ nm ($N=240$, $h=0.246$ nm), and temperatures are
$T_+=310$ K and $T_-=290$ K, the numbers of end segments interacting with
the thermostats $N_\pm=40$ (corresponding fragments are shown
in grey). Heat conductivity is $\kappa=177$ W/mK.
}
\label{fig3}
\end{figure}

\begin{figure}[tbh]
\includegraphics[angle=0, width=1.\linewidth]{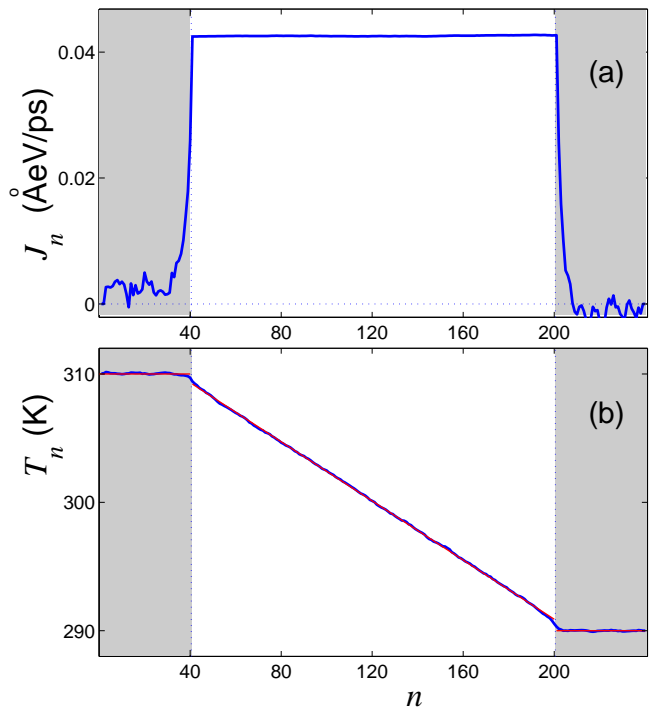}
\caption{
(Color online) Distribution of (a) local heat flux $J_n$  and (b) local average temperature
$T_n$ along a zigzag nanoribbon  with rough edges ($K=10$, density $d=0.87$).
Length of the nanoribbon is
$L=(N-1)h= 58.8$ nm ($N=240$, $h=0.246$ nm), and temperatures are
$T_+=310$ K and $T_-=290$ K.
%, the numbers of end segments interacting with
%the thermostats $N_\pm=40$ (corresponding fragments are shown
%in grey).
Heat conductivity is $\kappa=14$ W/mK.
}
\label{fig4}
\end{figure}

Distribution of the temperature and local heat flux along the rough-edged nanoribbon is shown
in Figs. \ref{fig3}(a,b) and \ref{fig4}(a,b).
The heat flux in each cross-section of the inner part of the nanoribbon
should remain constant, namely $J_n\equiv J$ for $N_+<n\le N-N_-$. The requirement of
independence of the heat flux $J_n$ on a local position $n$ is a good criterion for
the accuracy of numerical simulations, as well as it may be used to determine the
integration time for calculating the mean values of $J_n$ and $T_n$. As follows from the
figures, the heat flux remains constant along the central inner part of the nanoribbon.
%---------------------------- Fig. 3 ------------------------------------
%---------------------------- Fig. 3 ------------------------------------
%---------------------------- Fig. 4 ------------------------------------
%---------------------------- Fig. 4 ------------------------------------

A linear temperature gradient can be used to define the local coefficient of thermal
conductivity, $\kappa(N-N_+-N_-)=(N-N_--N_+-1)J/(T_{N_++1}-T_{N-N_-})S$,
where $S=2(dD_y+2r_C)r_C$ is the area of the nanoribbon cross-section
(nanoribbon width $D_y=(3K/4-1)\rho_0$, Van der Waals carbon radius $r_c=1.85$\AA).
Using this definition, we can calculate the asymptotic value of the coefficient
$\kappa=\lim_{N\rightarrow\infty} \kappa(N)$.

\section{Results}
Our analysis of linear eigenmodes of the nanoribbon with periodic
boundary conditions in $n$ reveals that in the case of edge disorder almost all
vibrational modes are localized as functions of the longitudinal index $n$. This means
that in our system we observe the manifestation of the Anderson localization
due to the edge disorder, earlier discussed only for the wave transmission in
surface-disordered waveguides~\cite{valya1,valya2}.

To analyze oscillation eigenmodes, we define the distribution
function of the oscillatory energy along the nanoribbon as follows:
$$
p_n=\sum_{k=1}^{K_n} M_{n,k}(|e_{n,k,1}|^2+|e_{n,k,2}|^2+|e_{n,k,3}|^2)/M_0,
$$
where $n=1,2,...,N$, $M_{n,k}$ is mass of the atom with the index $(n,k)$, and
$\{e_{n,k,i}\}_{i=1}^3$ is a component of the corresponding eigenvector (see Ref.~\cite{sk10prb}).
The energy distribution is normalized in accord with the condition:
$
\sum_{n=1}^N p_n=1.
$
To describe the longitudinal energy localization, we introduce 
a new parameter,
$
D=1/\sum_{n=1}^Np_n^2,
$
that characterizes the width of the energy
localization along the nanoribbon. If a vibrational mode is
localized only on one elementary cell, the corresponding width is $D=1$. In the opposite limit,
when the vibrational energy is distributed equally on all elementary cells,
we have $D=N$, so that in a general case $1\le D\le N$.

\begin{figure}[tb]
\includegraphics[angle=0, width=1.\linewidth]{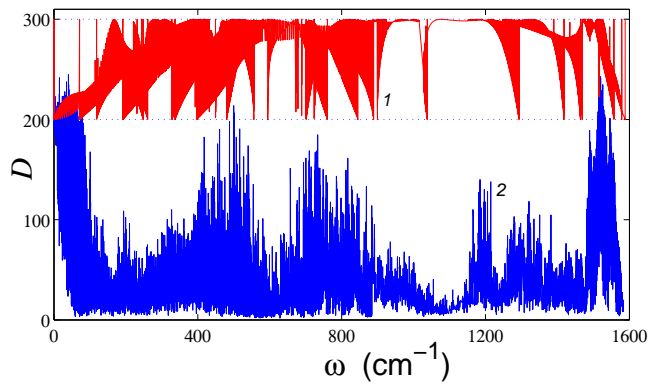}
\caption{
(Color online) Dependence of the width parameter $D$ of linear eigenmodes for the nanoribbon
with periodic boundary conditions ($n+N\equiv n$, length $N=300$) on the frequency
$\omega$ for an ideal nanoribbon (curve 1, width $K=10$) and a nanoribbon with rough edges
(curve 2, $K=10$, $p=0.5$).
}
\label{fig5}
\end{figure}

Dependence of the width $D$ on the frequency of the oscillatory eigenmodes
is shown in Fig. \ref{fig5}. For an ideal nanoribbon, all modes are not localized: when
the length $N=300$ we have the width $200\le D\le 300$. For nanoribbons with rough edges,
only the modes with the wavelength of the order of the nanoribbon length
are not localized. As a result, we expect that the edge disorder should lead to suppression 
of phonon transport and dramatic reduction of the thermal conductivity.

Our numerical results demonstrate that the thermal conductivity of graphene nanoribbon depends
crucially on the degree of edge roughness. In spite of the fact that the nanoribbon has
an ideal internal structure, its thermal conductivity is reduced dramatically, and it becomes
much lower that the conductivity of an ideal nanoribbon of the same width.
%---------------------------- Fig. 5 ------------------------------------
%---------------------------- Fig. 5 ------------------------------------
%---------------------------- Fig. 6 ------------------------------------
\begin{figure}[tb]
\includegraphics[angle=0, width=1.\linewidth]{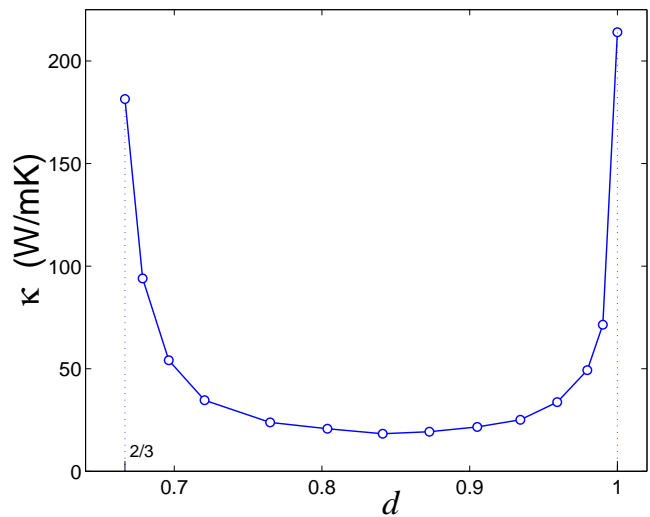}
\caption{
Dependence of the coefficient of thermal conductivity $\kappa$ of a finite nanoribbon
with rough edges ($N=240$, $N_\pm=40$, $K=12$) on the density $d$.
}
\label{fig6}
\end{figure}
%---------------------------- Fig. 6 ------------------------------------

Distribution of the thermal flow $J_n$ and local temperature $T_n$ along the
ideal nanoribbon and the nanoribbon
with rough edges (for the density $d=0.87$) are  presented in Figs.
\ref{fig3}(a,b) and \ref{fig4}(a,b).
In comparison with the ideal nanoribbon, the edge disorder leads to
reduction of the thermal flow in at least ten times,
as well as it changes the temperature profile along the nanoribbon.
In addition, in an ideal
nanoribbon we observe thermal resistance at the edges placed into a thermostat,
which disappears in the case of rough surfaces.
As a result, for the length $L=(N-N_--N_+)h=39.4$ nm, ($N=240$, $N_\pm=40$) the
coefficient of thermal conductivity of the nanoribbon with rough edges is found as
$\kappa=14$ W/mK that is in 12.6 times lower than the thermal conductivity of an ideal
nanoribbon, $\kappa=177$ W/mK.
%---------------------------- Fig. 7 ------------------------------------
\begin{figure}[tb]
\includegraphics[angle=0, width=1.\linewidth]{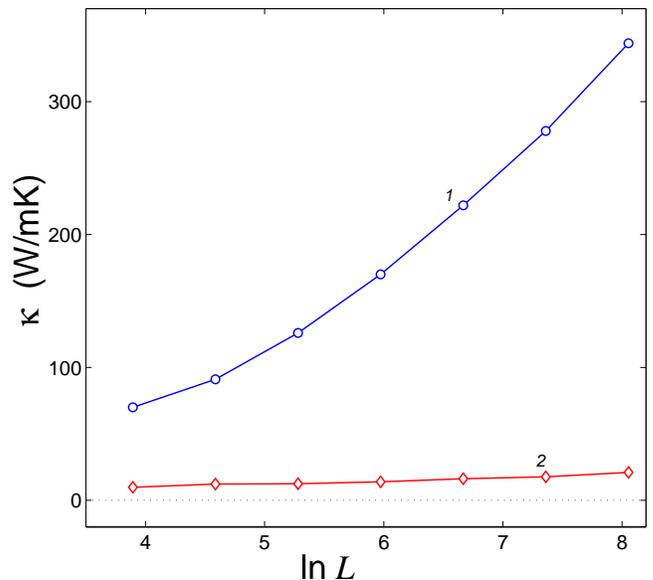}
\caption{
Dependence of the coefficient of thermal conductivity $\kappa$
on the length of the central part of the nanoribbon $L$ (dimension $[L]=$\AA).
Curve 1 corresponds to an ideal nanoribbon with $K=8$, curve 2 --
to edge-disordered nanoribbon with
$K=10$, $p=0.5$.
}
\label{fig7}
\end{figure}
%---------------------------- Fig. 7 ------------------------------------

Dependence of the coefficient of thermal conductivity $\kappa$ on the degree of roughness
characterized by the parameter $d$ is shown in Fig. \ref{fig6} for $K=12$, $N=240$, and $N_\pm=20$.
As follows from this figure, the thermal conductivity will be the lowest for the densities
$0.76\le d< 0.93$ (corresponding to the probability of
removing the edge atoms, $0.2<p<0.7$). The maximum is observed for  $d=1$ (probability $p=0$)
and $d=(K-4)/K=2/3$ ($p=1$) when we have ideal nanoribbons with $K=12$ and  $K=8$ atoms
in an elementary cell, respectively. The minimum is observed
for the density $d=0.84$ (probability $p=0.4$).
Below, we consider nanoribbons with rough edges created by removing edge atoms
with the probability $p=0.5$.
The corresponding structure of this nanoribbon is shown in Fig. \ref{fig2}.
%---------------------------- Fig. 8 ------------------------------------
\begin{figure}[t]
\includegraphics[angle=0, width=1.\linewidth]{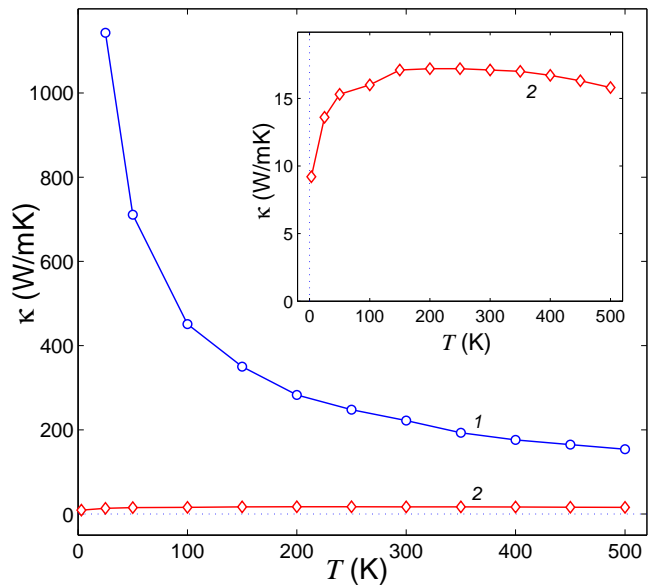}
\caption{
Dependence of the thermal conductivity coefficient $\kappa$
on the temperature $T$ for the ideal finite nanoribbon (curve 1, width $K=8$)
and for edge-disordered nanoribbon (curve 2, width $K=10$, probability $p=0.5$).
Length of the central part $L=320h=78.6$nm ($N=400$, $N_\pm=40$).
}
\label{fig8}
\end{figure}
%---------------------------- Fig. 8 ------------------------------------

Our numerical modeling described above demonstrates that for $T=300$K the thermal
conductivity of an ideal nanoribbon grows with its length $L$ as a power-law function,
$\kappa\sim L^\alpha$ for $L\longrightarrow\infty$
where $\alpha\approx 1/3$. In contrast, the thermal conductivity of a nanoribbon with
rough edges grows much slower, see Fig. \ref{fig7}. This difference grows with the length
of the nanoribbon. For example, for $L=4.91$ nm $(N=120$, $N_\pm=40$),
a ratio $\beta$ between the coefficient of thermal conductivity of
disordered ($K=12$, $p=0.5$) nanoribbon, $\kappa_1$, and ideal ($K=10$, $p=0$) nanoribbon,
$\kappa_0$, of the same width is $\beta=\kappa_1/\kappa_0=0.14$, but for the
length $L=314.4$ nm this ratio becomes much smaller, $\beta=0.06$.

Efficiency of the thermal conductivity of edge-disordered nanoribbons also decreases with
temperature, as well as with the ratio $\beta$.
For an ideal nanoribbon, the coefficient of thermal conductivity grows monotonically for low
temperatures (see Fig. \ref{fig8}, curve 1), so that for $T\rightarrow 0$ we obtain $\kappa\rightarrow\infty$.
This is related to the fact that the dynamics of nanoribbons approached the dynamics of
one-dimensional linear system with infinite thermal conductivity.
In contrast, for the nanoribbon with rough edges we observe that for $T>100$K its thermal
conductivity depends only weakly on temperature, see Fig. \ref{fig8}, curve 2.
This result is explained by the fact that in the edge-disordered nanoribbon
all linear vibrational modes becomes localized due to the edge disorder, and the phonon
transport is suppressed. For low temperatures,
the system become linear and its thermal conductivity decays, since a diffusion
transport is driven by nonlinear dynamics.
As a result, the ratio  $\beta=\kappa_1/\kappa_0$ decays monotonically.
For example, for the nanoribbon with the length $L=78.6$ nm, at $T=500$K this ratio is $\beta=0.10$,
at $T=300$K, we have $\beta=0.077$, at $T=100$K we obtain $\beta=0.036$, at $T=50$K we
find $\beta=0.021$, and at $T=25$K, we have $\beta=0.012$ (i.e. the thermal conductivity
is reduced by two orders!).

\section{Conclusions}

We have studied numerically thermal conductivity of carbon nanoribbons
with ideal and rough edges. We have demonstrated that thermal conductivity 
of an ideal nanoribbon is a monotonic power-like function of its length. However, 
the thermal conductivity is modified dramatically when the structure of nanoribbon edges change. 
In particular, the thermal conductivity of a nanoribbon with an edge-induced disorder is reduced 
by several orders of magnitude, and this effect is more pronounced
for longer ribbons and low temperatures. As a result, nanoribbons  with ideal edges can
play a role of highly efficient conductors, while nanoribbons with rough edges become
efficient thermal resistors.

\section*{Acknowledgements}

Alex Savin acknowledges a hospitality of the Center for Nonlinear Studies of the
Hong Kong Baptist University and Nonlinear Physics Center of the
Australian National University where this work has been completed. This work was
supported by the Australian Research Council.

\bibliography{nano}

\end{document}